\documentclass[aps,pra, onecolumn,14 pts,a4paper,showpacs]{revtex4}
\usepackage{times}
\usepackage{graphicx}
\usepackage{bm}
\usepackage{color}

\usepackage{color}

\definecolor{pink}{rgb}{1,1,0} 
\definecolor{red}{rgb}{1,0,0}
\definecolor{yellow}{rgb}{1,1,0}
\definecolor{orange}{rgb}{1,0.5,0}
\definecolor{white}{rgb}{1,1,1}

\begin{document}

\title{Foretelling catastrophes?}

\author{Yves Pomeau$^1$ and Martine Le Berre$^2$ }

\affiliation{$^1$Department of Mathematics, University of Arizona, Tucson, USA.
\\ $^2$ Institut des Sciences Moléculaires d'Orsay ISMO-CNRS, Univ. Paris-Sud, Bat. 210, 91405 Orsay, France.}

\begin{abstract}

A generic saddle-node bifurcation is proposed to modelize fast transitions of finite amplitude arising in geophysical (and perhaps other) contexts, when they result from the intrinsic dynamics of the system. The fast transition is generically preceded by a precursor phase which is less rapid, that we characterize. In this model, if an external source of noise exist, the correlation length of the fluctuations increases before the transition,  and its spectrum tends to drift towards lower frequencies.  This change in the fluctuations could be a way of detecting catastrophic events  before they happen.

\end{abstract}

\maketitle

\date{\today }

\section{Statement of the problem}
Earthquakes, like volcanic eruptions as well as other physical phenomena and perhaps also some kinds of socio-economical "revolutions", show an abrupt transition from one state to another. We consider cases where this transition is intrinsic (not the result of an excitation from outside) and dynamical in the sense that, as a parameter changes slowly, the system jumps by a finite amount in a time much shorter than the typical time of evolution of the external parameter. In earthquake physics, this typical time of evolution, the earthquake recurrence time, is on geological scales of  plate tectonics although the time scale of seismic ruptures is within the second to minute range \cite{Scho}.
Our basic assumption is that, as a dynamical system, an earthquake shows a "dynamical saddle-node" bifurcation. At the bifurcation point, a pair of fixed points, one  locally stable the other locally unstable, merge and vanish as a control parameter varies. Take a damped dynamical system, with a coordinate $x(t)$ solution of the equation
\begin{equation}
\frac{{\mathrm{d}}x}{{\mathrm{d}}t} = -\frac{\partial V}{\partial x}
\mathrm{.}
\label{eq:grad}
\end{equation}
In this equation $V(x)$ is a potential.
In the geophysical context of earthquakes, the scalar variable $x$ could be the relative displacement across the fault.
The equation (\ref{eq:grad}) is too general to be very helpful. However, as time varies slowly, it may describe a saddle-node bifurcation where a stable equilibrium disappears, assuming that $V$ depends slowly on time in a prescribed way, to become a function $V(x,t)$. Near the transition, one may use a
mathematical picture which is correct for a short time around the
transition if the potential $V(x,t)$ is a smooth function (see below for what happens beyond this local study).

Assume first that $V(.)$ does not depend explicitely on time and takes the form
\begin{equation}
 V(x) =- (\frac{1}{3} x^3 + b x)
 \mathrm{,}
 \label{eq:sd}
 \end{equation}
with $b$ real constant (for the moment).

\begin{figure}[htbp]
\centerline{$\;\;$
(a)\includegraphics[height=1.0in]{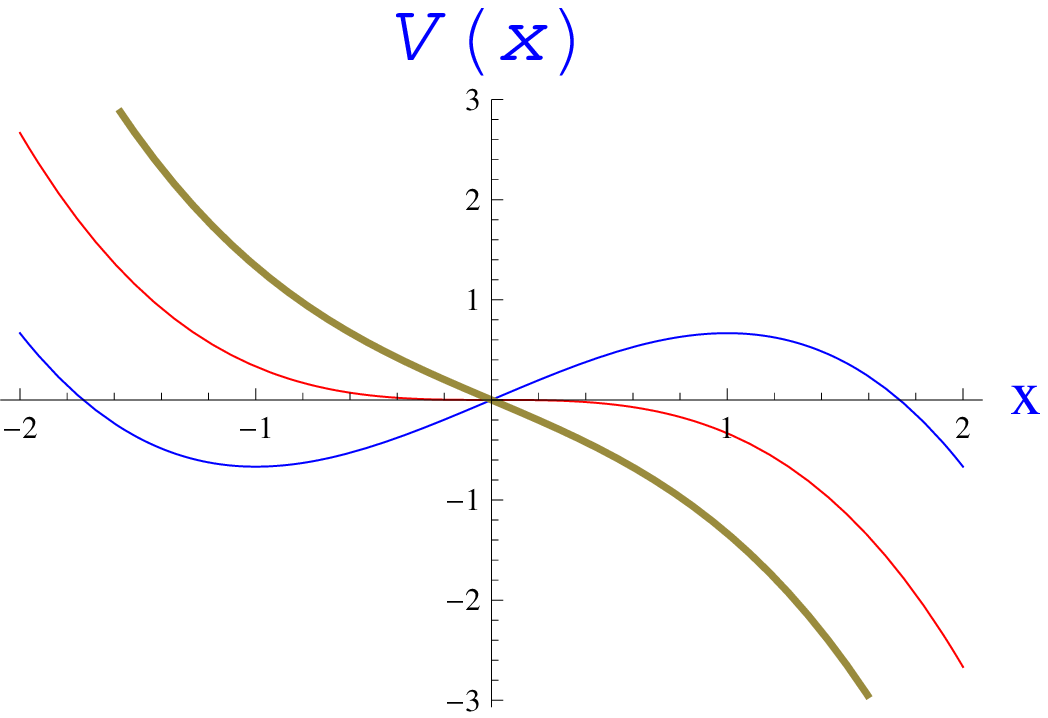}
(b)\includegraphics[height=1.0in]{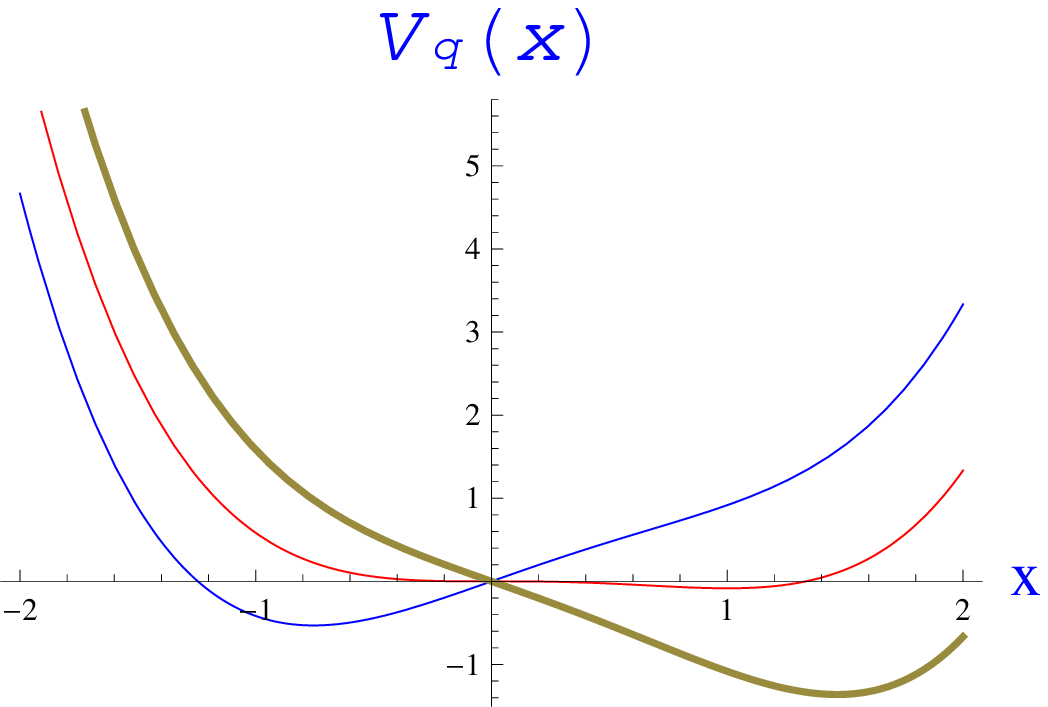}$\;\;$
}
\caption{(a) Cubic potential for $b =-1,0,1$. (b) Quartic potential,  $ b=-1,0,1$}
\label{Fig:pot}
\end{figure}

For $b$ negative $V(x)$ has two real extrema ({\it{i.e.}} the roots of $\frac{\partial V}{\partial x} =0$), one $-\sqrt{-b}$ is a stable equilibrium, the other, $\sqrt{-b}$, is an unstable equilibrium.
For $b =0$ the two equilibria merge and disappear for $b$ positive, see Figure (\ref{Fig:pot}-a). This is the saddle-node bifurcation. The shape of $V(x)$ near $x =0$ and for $b$ small is universal: for a given smooth $V(x)$ showing this saddle-node bifurcation, one can always rescale time and external parameter to find the "local' problem in this form.

The extension to time dependent variation of the control parameter $b$ goes as follows. If $b$ is a smooth function of time, one can assume that $b(t)$ crosses the critical value, {\it{i.e.}} zero in the present case, at time zero in such a way that $ b(t) = a t +...$ with $a$ non zero constant, and the dots being for higher terms in the Taylor expansion of $b(t)$. For $t$ and $x$ close to zero, after elementary rescaling, one can represent
 the dynamical system (\ref{eq:grad}), close to the saddle-node bifurcation, by an "universal"
parameterless equation

\begin{equation}
\frac{{\mathrm{d}}x}{{\mathrm{d}}t} = x^2 + t
\mathrm{.}
\label{eq:gradt}
\end{equation}

Outside of the neighborhood of $x=0$, the solution of (\ref{eq:grad}) depends on other parameters like the one defining $V(.)$ far from $x =0$ , as studied
below.
\subsection{Solution close to the saddle-node}
\label{sec:close}

Let first consider what happens close to the saddle-node bifurcation. We look for a solution of equation (\ref{eq:gradt}) transiting from the "stable" fixed point at "large" negative times to the rolling down one towards positive value of $x$ at positive times. This  solution behaves like $x(t) \approx - \sqrt{-t}$ at large negative times. The equation (\ref{eq:gradt}) is of the Riccatti type and can be integrated by introducing the function $y(t)$ such that $x(t) = -\frac{y'}{y}$ where $y' = \frac{{\mathrm{d}}y}{{\mathrm{d}}t}$ and $y(t)$  is a solution of Airy's equation,

\begin{equation}
y'' + t y = 0
\mathrm{.}
\label{eq:gradA}
\end{equation}

\begin{figure}[htbp]
\centerline{$\;\;$
(a)\includegraphics[height=1.5in]{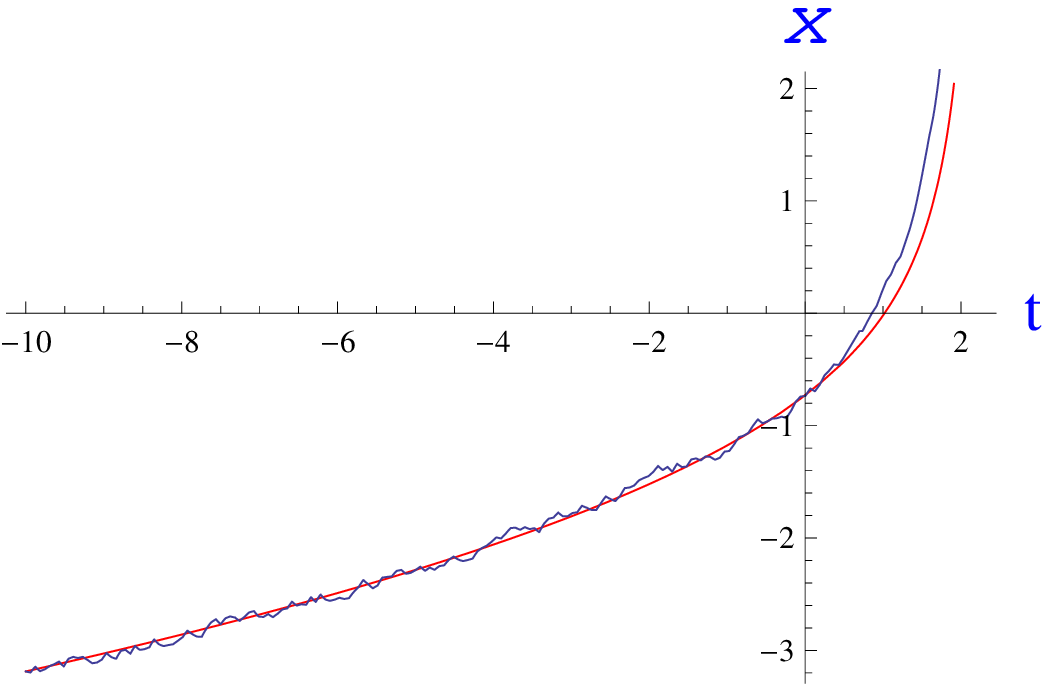}
(b)\includegraphics[height=1.in]{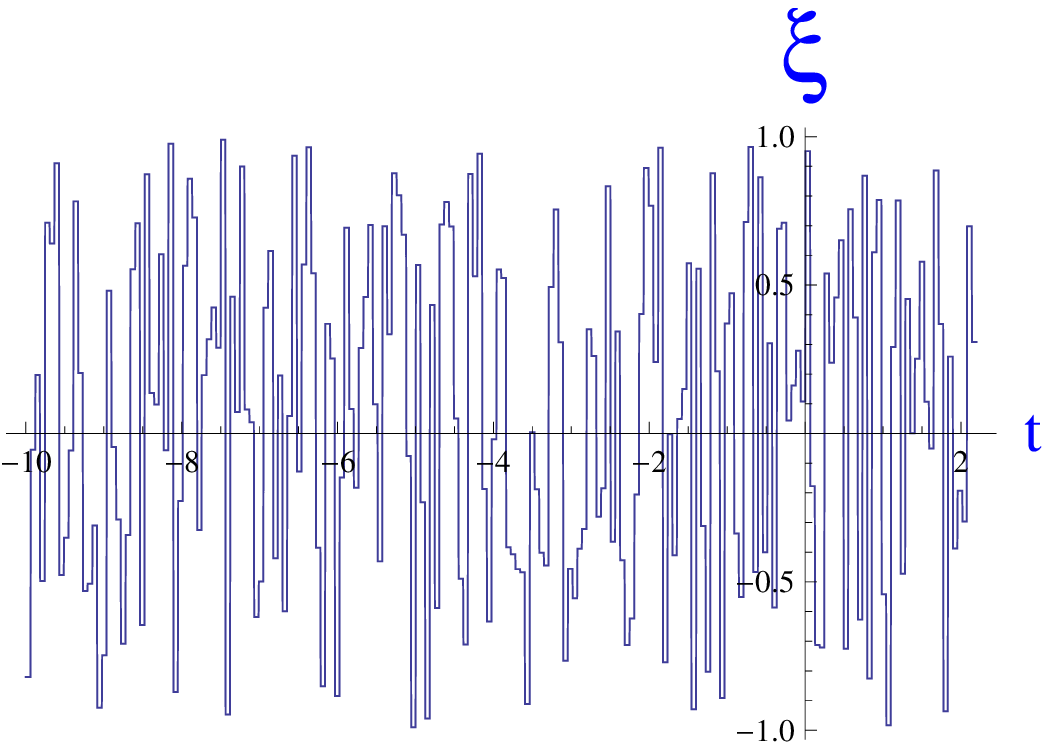}
(c) \includegraphics[height=1.5in]{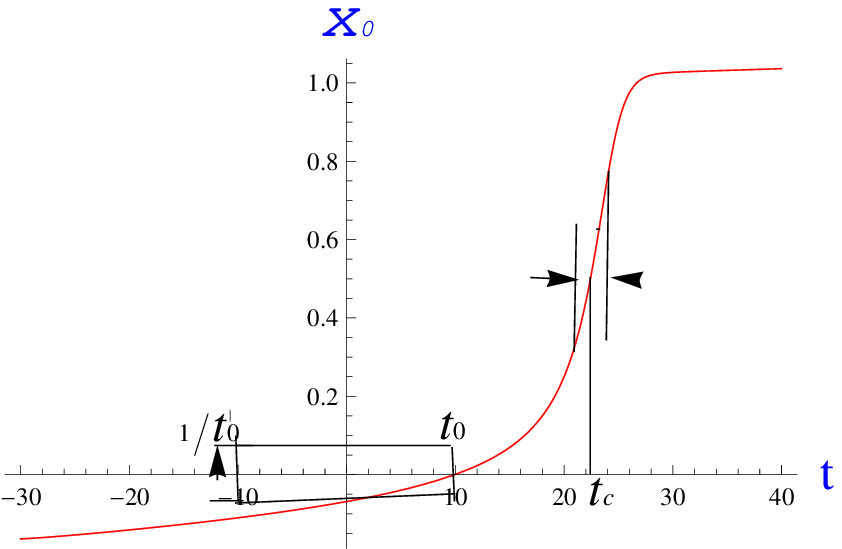}
$\;\;$}
\caption{(a) Solutions of equation (\ref{eq:gradtn}), with and without noise,
$\epsilon=0$ (smooth curve) and $\epsilon=1$ (noisy curve). (b) Noise $\zeta(t)$. (c) Solution of equation (\ref{eq:dynquarticscaling})  for $a=10^{-3}$. The rectangle around the origin defines the region  $-t_0 < t <t_0$  and $-1/t_0 < x <1/t_0$, with $t_0\sim a^{-1/3}$. The critical time is $t_c\sim 2.34$ $ t_0$. The two vertical lines inserted between the two arrows delimitate the large slope time duration, of order unity.}
\label{Fig:cubic}
\end{figure}

The solution relevant with the given condition for $x(t)$ at $t\rightarrow-\infty$, is the Airy function  $Ai(-t)$  which writes
\begin{equation}
 Y(t) = Ai(-t)=\int_0^{+\infty} \cos(\frac{u^3}{3} - ut) {\mathrm{d}}u
 \mathrm{,}
 \label{eq:Y}
\end{equation}
 and leads to the curve $x(t)$  drawn on Figure(\ref{Fig:cubic}-a).
Yet we have only solved the transient problem near the saddle-node bifurcation.  The transition ends-up when $t$ becomes equal to the first zero of the Airy function $Ai(-t)$, corresponding to a divergence of the original $x(t)$. Let
$t_c$ be this critical value of $t$, {\it{i.e.}} the smallest root of
$Y(t) = 0$, a pure number, about  $t_c \approx 2.338$, and let us look at the
behaviour of $x(t)$ just before this transition. From the Laurent expansion of $Y(t)$ close to $t_c$, and returning to the original variable $x(t)$,
one obtains
\begin{equation}
x(t) \approx \frac{1}{t_c - t} -\frac{t_c}{3}(t_c-t)...
\mathrm{.}
 \label{eq:x0exp}
 \end{equation}

As this solution diverges, it looses its validity because the "universal" dynamical equation  (\ref{eq:gradt}) was derived under the condition that $x$ remains close to zero. This local theory cannot deal with finite variations
away from the critical conditions, therefore we shall add finite amplitude effects to limit the growth of the instability after the
transition.

We shall study now two questions,
first the response of this dynamical system to an external noise source,
then the dynamics of a system showing a saddle-node bifurcation of the
type just studied and reaching a new stable fixed point after this
bifurcation.
\subsection{Response to an external noise source}
\label{sec:response}

We explore first the response of our system to a small external noise, and search whether
the response to the noise
changes qualitatively and so could be a signal ahead of the transition.

Let us consider the equation (\ref{eq:gradt}) with a small noise
added, so that equation (\ref{eq:gradt}) is replaced
by

\begin{equation}
\frac{{\mathrm{d}}x}{{\mathrm{d}}t} = x^2 + t + \epsilon \zeta(t)
\mathrm{,}
\label{eq:gradtn}
\end{equation}
where $\zeta(t)$ is a random function of time, and $\epsilon$ a small factor.

In the limit  $\epsilon$ small, one can solve equation (\ref{eq:gradtn}) by expansion in powers of $\epsilon$,
\begin{equation}
 x(t) = x_0 (t) + \epsilon x_1(t) + ...
\mathrm{.}
\label{eq:serie}
\end{equation}

where
\begin{equation}
 x_0(t) =  -\frac{Y'(t)}{Y(t)}
\mathrm{,}
\label{eq:x0}
\end{equation}

The linear response to the noise is
\begin{equation}
 x_1 (t) = \frac{1}{Y^2(t)} \int_{t_0}^t {\mathrm{d}}\tilde{t} \ \zeta(\tilde{t}) \ Y^2(\tilde{t})
  \mathrm{.}
 \label{eq:x1.1}
 \end{equation}
Because $Y^2(\tilde{t})$ tends rapidly to zero as $(\tilde{t})$ tends to minus infinity, one can take $t_0 = -\infty$
to get rid of the effect of the initial conditions.


Let us take a delta-correlated (or white) noise, such that
\begin{equation}
\langle  \zeta(t_a) \zeta(t_b) \rangle = \delta(t_a - t_b)
\mathrm{.}
\label{eq:noise}
 \end{equation}
Note that, including with $\epsilon=1$, the solution of equation (\ref{eq:gradtn}) with noise is still very close to the noiseless solution $x_0(t)$.

The correlation function of $x_1(t)$ is given by
\begin{equation}
 \langle x_1 (t) x_1(t') \rangle  =
  \frac{1}{Y^2(t) Y^2(t')}  \int_{-\infty}^{\inf(t,t')} {\mathrm{d}}\tilde{t} Y^4(\tilde{t})
  \mathrm{,}
\label{eq:corr}
\end{equation}
whose behavior
for large negative values of both $t$ and $t'$,
is derived from the asymptotic expression of the Airy function, $Ai(-t) \approx \frac{e^{ - \frac{2}{3} (-t)^{3/2}}}{2\sqrt{\pi}(-t)^{1/4}}$.
Setting $w=\frac{\tilde{t}}{t}$, and $F(w)=1-w^{3/2}$, the variance $\sigma^2=\langle (x(t)-x_0(t))^2 \rangle$ of the fluctuations writes
\begin{equation}
\langle x_1 (t)^2 \rangle  \approx  (-t)  \int_{1}^{\infty} \frac{{\mathrm{d}}w }{w}e^{\frac{8}{3} (-t)^{3/2} F(w)}
 \mathrm{.}
 \label{eq:autocorr}
\end{equation}
In the limit $(-t) \rightarrow \infty$ the integral is concentrated near $w = 1$
so that

\begin{equation}
 \langle x_1 (t)^2 \rangle  \approx  \frac{1}{4}(-t)^{-\frac{1}{2}}
 \mathrm{,}
 \label{eq:limcor}
 \end{equation}

\noindent
which shows that the fluctuations increase as time goes on, some time before the transition itself.
As the transition approaches,
 the variance of the fluctuations
  increases close to the critical time $t_c$, because $Y(t_c)=0$.

    Because of the divergence of the solution  at $t=t_c$, it does not make sense to describe the dynamical behavior of the fluctuations due to the external noise very close to $t_c$.
This unbounded growth is a consequence of the {\emph{local}} cubic form of $V(x)$ as expanded
near $x =0$, which is valid around $x=0$ only,
 in obvious contradiction with the fact that $x(t)$ diverges.
 
 \section{Dynamical saddle-node bifurcation toward a new steady state}
\label{sec:dynamicalsaddle}
To suppress the divergence of $x(t)$ after
the saddle-node bifurcation we add a stabilizing (positive) term to the potential $V(x)$ which becomes quartic,

\begin{equation}
V_q(x) = - \frac{x^3}{3} - b x + \frac{x^4}{4}
\mathrm{,}
\label{eq:quartic}
\end{equation}

as drawn in Figure(\ref{Fig:pot}-b).
Because of the growth  of $V_q(x)$ at infinity, like $x^4$, the solution of the differential
equation
\begin{equation}
\frac{{\mathrm{d}}x}{{\mathrm{d}}t} = -\frac{\partial V_q}{\partial x} =b + x^2 - x^3,
\mathrm{.}
\label{eq:dynquartic}
\end{equation}
does not diverge at finite time.
The equation (\ref{eq:quartic}) can be written in the given scaled form,
provided the coefficient of $x^4$  is positive. For such a potential
one parameter only remains. In equation
(\ref{eq:quartic}) the
coefficient is chosen as $b$, the one of the linear term. For $b = 0$ the dynamical system
(\ref{eq:dynquartic})
is exactly at the saddle-node bifurcation, because at $b = x =0$ the first and second derivative of $V_q(x)$ both vanish, but not the third one.
Contrary to the case of the pure
cubic potential, this system has always, that is for any value of $b$, a
stable fixed point beyond the pair of fixed points collapsing at the
saddle-node bifurcation. This makes it a fair candidate for describing the
dynamical saddle-node bifurcation without blow-up.

 As in the previous case, we shall look now at the case of a time dependent $b$, that will be
 taken as $ b = a t$ with $a$ positive constant.
 Because of the rescaling of the cubic and quartic term, the parameter $a$ cannot be eliminated
  (another possibility is to put a parameter in front of the cubic term).
For the potential $V_q(x) = -\frac{x^3}{3} - a t x + \frac{x^4}{4}$ we shall analyse the solution of the dynamical equation
\begin{equation}
\frac{{\mathrm{d}}x}{{\mathrm{d}}t} = a t + x^2 - x^3
\mathrm{,}
\label{eq:dynquarticscaling}
\end{equation}
tending at large times to the quasi-equilibrium point $ x = (a t)^{1/3}$, $t$ being
considered as a parameter, see Figure(\ref{Fig:cubic}-c). Moreover we consider the limit $a$  small,
which could describe a wide range of slip phenomena \cite{slow-slip}, as earthquakes where $a$ is generally very small, of order $10^{-9}$ (see below).

\subsection{Three time ranges for small $a$}
\label{sec:a small}
 In this limit we show first that there are three characteristic time intervals, depending how $x$ is
close to zero.

The long time scale is the average recurrence time of an earthquake at the same site along the fault. It is typically of order $t_{b}^{{\rm phys}}\sim 200$ years. In our model it is the time needed for the potential $V_q(x,t)$ to
change significantly, to move from a pair of fixed points to a saddle-node
bifurcation. Because time enters in $V_q(x,t)$ through the combination
$(at)$, the adimensional time needed for a change of shape of $V_q$ is of order
\begin{equation}
t_b \sim a^{-1}
\mathrm{.}
\label{eq:temps}
\end{equation}

The short time $t_{{\rm eqk}}$ is of order unity in our model equation (\ref{eq:dynquarticscaling}) as stated in the next paragraph. It is the duration of the abrupt change of the slope of the solution $x(t)$. This short time corresponds to the dynamic rupture duration, which is typically of order ten seconds for a magnitude 6 earthquake, $t_{{\rm eqk}}^{{\rm phys}} \sim 10s$ \cite{Scho}. Therefore the ratio of these two time scales $\frac{t_{{\rm eqk}}^{{\rm phys}}}{t_b^{{\rm phys}}}=\frac{t_{{\rm eqk}}}{t_b}$, is small as $a=1.6$ $10^{-9}$ in the geophysical context.

There is another time scale, $t_0$, the time interval standing before the transition, and close to it, during which the potential is very flat, while the solution  has not jumped. During this time, $x$ and $at$  are much smaller than unity, then the cubic term on the right-hand side of equation
(\ref{eq:dynquarticscaling}) is negligible. In this range one recovers the universal
equation of the dynamical saddle-node bifurcation (\ref{eq:gradt}) by
taking $X = x a^{-1/3}$ and $T = t a^{1/3}$, with the boundary condition
$X(t) \approx - \sqrt{-T}$ at $T$ tending to minus infinity. This property concerns the rectangular domain drawn on Figure (\ref{Fig:cubic}-c), where $x$ is small, $x \sim a^{1/3}$, and  $t$ extends from $-a^{-1/3}$ to $t \sim
a^{-1/3}$, located before the abrupt increase. Therefore the time extension of this domain introduces the intermediate time scale,
\begin{equation}
t_0  \sim a^{-1/3}
\mathrm{,}
\label{eq:temps}
\end{equation}
long compared  to unity (the adimensional time scale $t_{{\rm eqk}}$ for the duration of a seismic rupture) and small compared to $t_b=a^{-1}$, the average recurrence time between earthquakes.

Let us prove that the short time is of order unity, by matching the solution $X(T)$ of
the universal equation
to the solution of equation (\ref{eq:dynquarticlate}) below, in the vicinity of the critical point $t_c(a)=a^{-1/3}$ $t_c$. Because $X(T)$ behaves like$\frac{1}{t_c-T}$ before it diverges, it follows that the solution
  $x(t)$ behaves as
$\approx \frac{1}{a^{-1/3}t_c- t}$ for "large" values of $\delta
t = t - a^{-1/3}t_c$  before the critical time. Using $\delta t$  in this
development as time variable,
$x(\delta t)$ becomes of order one when $\delta
t$ becomes of order one too. When this happens, the
term $at$ in equation
(\ref{eq:dynquarticscaling}) is negligible, therefore the solution of this
equation which can be matched with the solution near the bifurcation is
the solution of the integrable equation
\begin{equation}
\frac{{\mathrm{d}}x(\delta t)}{{\mathrm{d}} (\delta t)} =  x(\delta t)^2 - x(\delta t)^3
\mathrm{,}
\label{eq:dynquarticlate}
\end{equation}
 with the asymptotic behavior for very large negative times $x(\delta t)
\sim - \frac{1}{\delta t}$.
This equation shows that  the time scale for the earthquake rupture is of order one, because it has no explicit dependence with respect to the small parameter $a$. This result is confirmed by the numerics: For $a$ small we find that the rising time of $x(t)$ close to $t_c(a)$ (defined as the half-width of the slope $\frac{dx}{dt}$ solution of equation (\ref{eq:dynquarticscaling})) is $t_{{\rm eqk}} \sim 2.5$, independent of $a$. From the observational point of view, the catastrophe takes place during this time $t_{{\rm eqk}}$ of order one, because the displacement is of order one then, compared to the displacement of order $a^{1/3}$ taking place during time $a^{-1/3}$ typical of the  "universal" transition process.
The two solutions match in the range $1 \ll (-\delta t) \ll a^{-1/3}$. Supposing that the physical fast time scale for earthquakes is $t_{{\rm eqk}}^{{\rm phys}} \sim 10 s$, the intermediate time scale is $t_0^{{\rm phys}} \sim a^{-1/3}t_{{\rm eqk}}^{{\rm phys}}$, which is  a few hours for $a=10^{-9}$.

 \subsection{Precursor effects due to an external noise}
\label{sec:precursor}

 With a noise source added, the dynamical equation (\ref{eq:grad})  becomes,
\begin{equation}
\frac{{\mathrm{d}}x}{{\mathrm{d}}t} = x^2 - x^3  + a t +\epsilon \zeta(t)
\mathrm{.}
\label{eq:xxquartic}
\end{equation}

Actually the effective noise amplitude is not equal to $\epsilon$ close to the saddle-node, but depends on the value of the parameter $a$. Indeed for $|t|\leq t_{0}$ , the cubic term in equation (\ref{eq:xxquartic}) is negligible, and the equation reduces to
\begin{equation}
\frac{{\mathrm{d}}x}{{\mathrm{d}}t} = x^2 + a t +\epsilon \zeta(t)
\mathrm{.}
\label{eq:xxquartic}
\end{equation}
which may be written on a form  $\frac{{\mathrm{d}}X}{{\mathrm{d}}T} = X^2 +  T +\tilde{\epsilon}(a) \zeta(t)$,
by setting $X=x a^{-1/3}$,  $T=ta^{1/3}$, and $\tilde{\epsilon}(a)=\epsilon a^{-2/3}$.
Therefore the effective noise is larger than $\epsilon$  in the rectangular domain of figure (\ref{Fig:cubic}-c).


Let us study the fluctuations of the solution $x(t)$ of equation (\ref{eq:xxquartic}).
For a small noise input, the solution may be expanded in power of $\epsilon$ as above. At first order it gives
\begin{equation}
\frac{{\mathrm{d}}x_1}{{\mathrm{d}}t} = [2x_0(t) -3 x_0^2(t)] x_1(t) + \zeta(t)
\mathrm{,}
\label{eq:x1}
\end{equation}
whose solution is formally
\begin{equation}
x_1 (t) = \int_{t_0}^t {\mathrm{d}}\tilde{t} \ \zeta(\tilde{t}) \exp[g(t)-g(\tilde{t})]
\mathrm{,}
\label{eq:x1i}
\end{equation}
where $g(t)$ is the time integral of the second derivative of the potential $-\frac{d^2V_q(x)}{dx^2}$, $g(t)=\int_{t_0}^t [2x_0(u) -3 x_0^2(u)]$.
The standard deviation $\sigma_{x_1}(t)$ has to be calculated numerically. We expect it to display the same behavior as for the cubic case in the whole domain where $x(t) \ll 1$ , i.e. before the transition, and close to it, because the potential is cubic in this region. After the transition, we expect that the fluctuation decreases, because the solution without noise becomes quasi-steady. This is confirmed by the numerics: as for the cubic potential, the fluctuations strongly increase close to the critical time $t_c$. With respect to time, the maximum of $\sigma_{x_1}$ occurs at time $t_c(a)$ for small noise. Therefore  the variance of the signal fluctuations cannot be used as a precursor for predicting the transition. Such a correlation between the standard deviation of the fluctuations and the sudden change of the solution  has been reported recently \cite{allen}  where the GPS geodetic signal, which can be assimilated to our $x_0(t)$, is shown to be strongly correlated to the seismic signal (which we see as related to the fluctuations $x_1(t)$).
Note that when the noise increases, the growth of the fluctuations occurs earlier and earlier, their maximum progressively shifting before $t_c$. This shift becomes visible only for an  effective noise amplitude larger than unity, that is physically outside the range of noise values.

Consider the case of small effective noise, where the correlation function and the spectrum of the fluctuations $x(t)-x_0(t)$ are well described by the correlation function and spectrum of $x_1(t)$, respectively. The calculation of these functions  requires some care because the system is not in a statistically steady state. Therefore the
spectral density of the fluctuations depends on time and the correlation function
\begin{equation}
\Gamma _{x_1}(t,\tau)= \langle x_1 (t-\tau/2) x_1(t+\tau/2) \rangle
 \mathrm{,}
 \label{eq:gamma}
 \end{equation}
 depends both on $t$ and on $\tau$.  A time dependent spectrum is formally defined by the (real) Wigner transform
 \begin{equation}
\mathrm{S} _{x_1}(t,f)=   \int_{-\infty}^{-\infty} {\mathrm{d}}\tau e^{-2i\pi f \tau} \langle x_1 (t-\tau/2) x_1(t+\tau/2)\rangle
 \mathrm{,}
 \label{eq:spec}
 \end{equation}
that has to be modified for numerical applications, by introducing a slipping window.

 \begin{figure}[htbp]
\centerline{$\;\;$
(a)\includegraphics[height=1.5in]{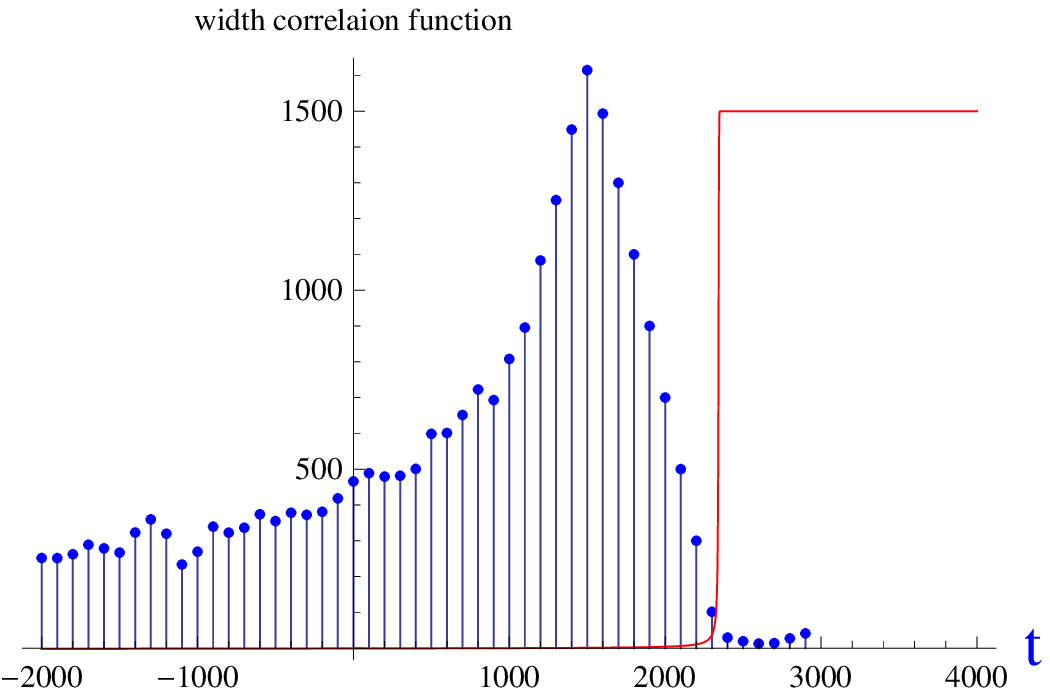}
(b)\includegraphics[height=1.5in]{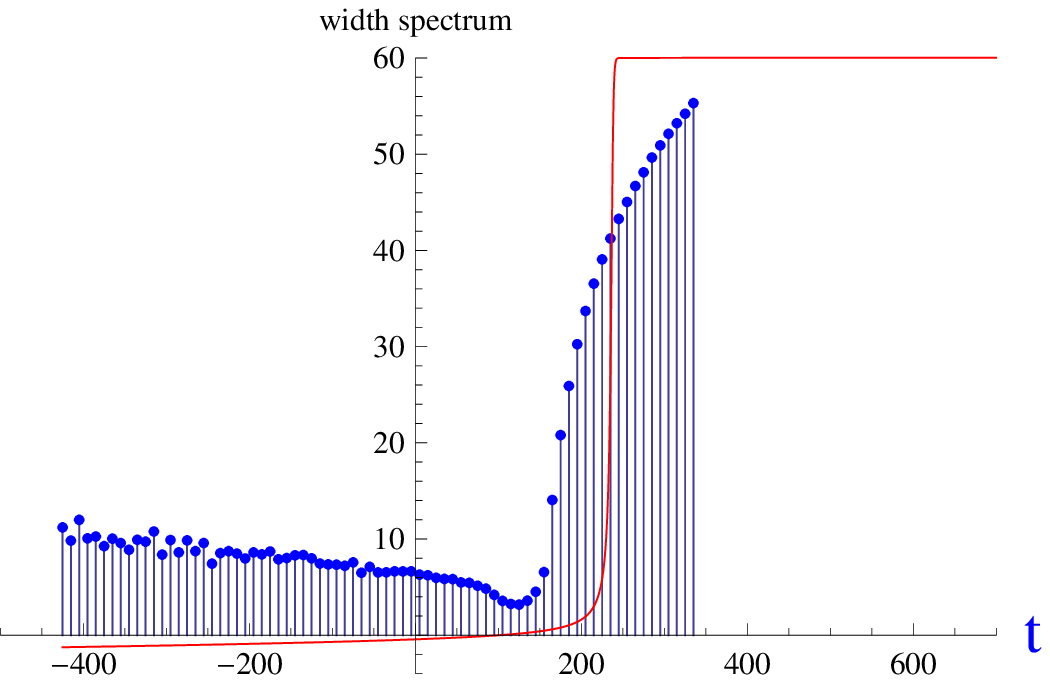}
$\;\;$}
\caption{(a) Width (in arbitrary units) of the correlation function of the fluctuation $x(t)-x_0(t)$ for $a=10^{-9}$ ($t_0=10^3$);
 (b) Spectral width (a. u.)}
\label{Fig:width}
\end{figure}
\begin{figure}[htbp]
\centerline{$\;\;$
(a)\includegraphics[height=1.5in]{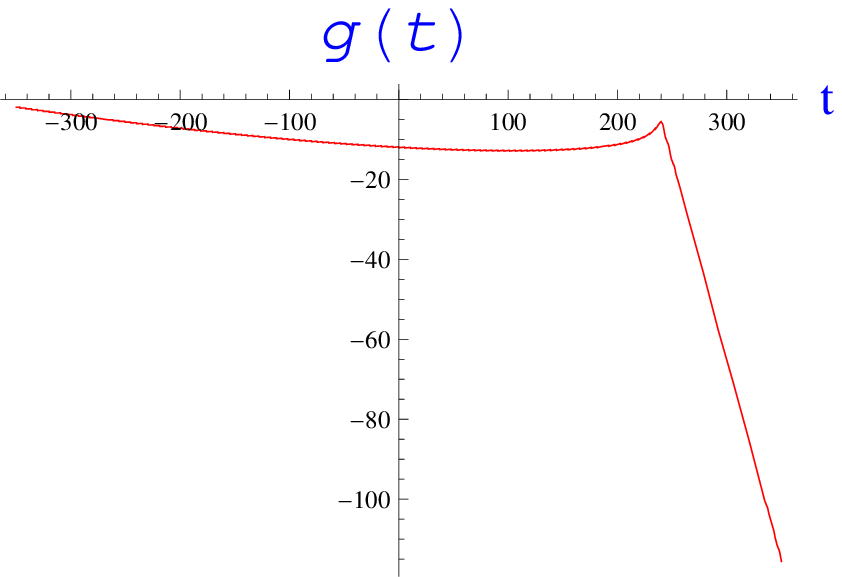}
(b)\includegraphics[height=1.5in]{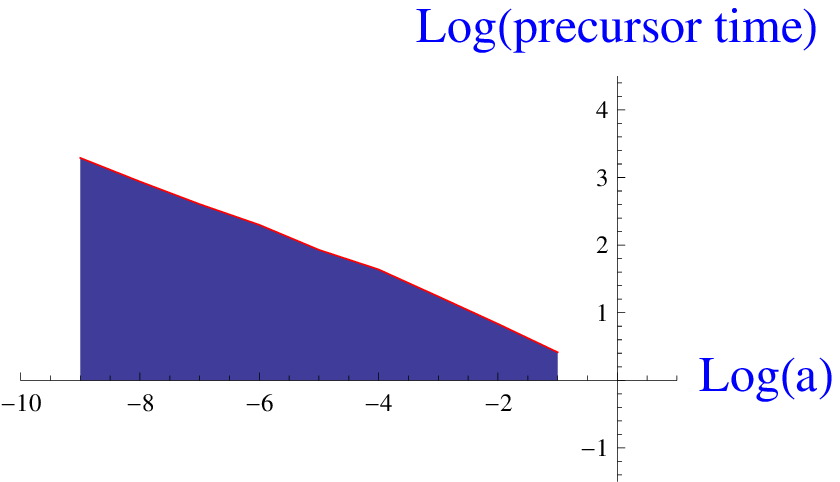}
$\;\;$}
\caption{(a) Flat shape of $g(t)$ before the critical time, for $a=10^{-9}$, or $t_0=1000$ ; (b) "Precursor time" $t_{prec}=t_c(a)-t_{1/3}$ versus $a$, 
in Log scale ( $t_{1/3}$ being the time where the width is $\frac{1}{3}$ of its maximum value, before the catastrophe time $t_c(a)$).}
\label{Fig:plat}
\end{figure}

 The width $\tau_x$ of the correlation function and the spectral width $\Delta f$ are reported in figures (\ref{Fig:width}) in a range of time of few $t_0$ around the transition, together with the solution ${x}_0(t)$ drawn in solid red line for covering.  Both widths show an interesting behavior which provides the same result. Consider first the left curve, obtained for the parameter value  $a=10^{-9}$, typical for earthquake phenomena, as discussed above.
  The figure displays a strong increase of the correlation time $\tau_x$ of $x(t)-x_0(t)$  in the intermediate time range, reaching its maximum value at time $t\sim 1.5$ $t_0$ (which was estimated as a couple of hours before the earthquake), then it displays a rapid decrease before the critical time $t_c(a) \sim 2340$. In addition, we observe a slow growth of $\tau_x$  as $t$ increases from large negative values (not shown in the figure), $\tau_x$ increasing by a factor ten for $-100t_0<t<0$, that corresponds to a time interval about one week.
    Such a remarkable behavior should be used as precursor.
 The increase of the correlation length \textit{before} the catastrophe can be understood when looking at the formal expression (\ref{eq:x1i}). The second derivative of the potential vanishes at $t=t_0$, that leads to the flatness of $g(t)$ in the whole domain $0 <t <t_c(a)$, as shown in figure (\ref{Fig:plat}-a). Figure (\ref{Fig:plat}-b) shows that the "precursor time" (see caption) is given by the relation
 
  \begin{equation}
t_{prec} \sim 2 t_0
 \label{eq:prec}
 \end{equation}
 which corresponds to about $4$ hours.
 
 As for the spectral width, the result is just the opposite: it continuously decreases from  large negative time, until the time $t\sim 1.5$$t_0$, where it suddenly grows. The slow decrease of $\Delta f$ corresponding to a slow shift of the spectrum  towards low frequencies, is followed by a rapid spectral broadening at the end of the intermediate domain, before the transition time $t_c(a)$. The two stages of the width change are both important, because they occur \textit{before the transition}.

  The growth of the fluctuations and their shift to lower frequencies can be understood as follows. As the transition approaches the potential $V(x,t)$ becomes flatter and flatter, making weaker and weaker the restoring force toward the equilibrium. Therefore, at constant noise source, the amplitude of the fluctuations driven by this noise source will grow because the damping is less and less efficient. Moreover, the typical time scale for this damping will get larger and larger because of the decreasing stiffness of the potential, which will favour noise at lower and lower frequencies.

\begin{acknowledgments}
 Jean-Louis LeMou\"{e}l and Cl\'{e}ment Narteau are
gratefully acknowledged for stimulating discussions, and Patrice Fromy for his technical help.

\end{acknowledgments}

	  \end{document}